\documentclass[10pt,conference]{IEEEtran}

\IEEEoverridecommandlockouts

\usepackage{cite}
\usepackage{flushend}
\usepackage{balance}

\usepackage{amsmath,amssymb,amsfonts}

\usepackage{graphicx}

\usepackage{multirow}

\usepackage{xurl}

\usepackage[table,xcdraw]{xcolor}
\usepackage[many]{tcolorbox}

\usepackage{fontawesome}

\usepackage{hyperref}

\usepackage{booktabs}

 \usepackage{todonotes}
\newcommand{\ReviewerC}[1]{\textcolor{black}{#1}}

\definecolor{main}{HTML}{CFCFCF}  
\definecolor{sub}{HTML}{CFCFCF}

\tcbset{
    sharp corners,           
    breakable = true,
    colback = white,         
    before skip = 0.2cm,     
    after skip = 0.5cm       
}

\newtcolorbox{boxC}{
    colback = sub,  
    boxrule = 0pt   
}

\newcounter{keyTakeAwaysCounter} 
\def\BibTeX{{\rm B\kern-.05em{\sc i\kern-.025em b}\kern-.08em
    T\kern-.1667em\lower.7ex\hbox{E}\kern-.125emX}}

\AtBeginDocument{
  \providecommand\BibTeX{{
    Bib\TeX}}}

\begin{document}

\title{Evaluating Large Language Models for Detecting Architectural Decision Violations}

 \author{
\IEEEauthorblockN{Ruoyu Su\IEEEauthorrefmark{1}, Alexander Bakhtin\IEEEauthorrefmark{1}, Noman Ahmad\IEEEauthorrefmark{1},  Matteo Esposito\IEEEauthorrefmark{1},Valentina Lenarduzzi\IEEEauthorrefmark{2}\IEEEauthorrefmark{1}, Davide Taibi\IEEEauthorrefmark{2}\IEEEauthorrefmark{1} }
\IEEEauthorblockA{\textit{\IEEEauthorrefmark{1}University of Oulu, Finland \quad
\IEEEauthorrefmark{2}University of Southern Denmark, Vejle, Denmark}}
\IEEEauthorblockA{\{ruoyu.su, alexander.bakthin, noman.ahmad, matteo.esposito\}@oulu.fi \quad \{lenarduzzi,taibi \}@imada.sdu.dk}
}

\maketitle

\begin{abstract} Architectural Decision Records (ADRs) play a central role in maintaining software architecture quality, yet many decision violations go unnoticed because projects lack both systematic documentation and automated detection mechanisms. Recent advances in Large Language Models (LLMs) open up new possibilities for automating architectural reasoning at scale.
We investigated how effectively LLMs can identify decision violations in open-source systems by examining their agreement, accuracy, and inherent limitations. Our study analyzed 980 ADRs across 109 GitHub repositories using a multi-model pipeline in which one LLM primary screens potential decision violations, and three additional LLMs independently validate the reasoning. We assessed agreement, accuracy, precision, and recall, and complemented the quantitative findings with expert evaluation.
The models achieved substantial agreement and strong accuracy for explicit, code-inferable decisions. Accuracy falls short for implicit or deployment-oriented decisions that depend on deployment configuration or organizational knowledge.
Therefore, LLMs can meaningfully support validation of architectural decision compliance; however, they are not yet replacing human expertise for decisions not focused on code. 
\end{abstract}


\section{Introduction}
Architectural Decision Records (ADR) are essential for preserving the integrity and long-term maintainability of software systems \cite{bosch2004software,buchgeher2023using}. They capture the reasoning behind architectural choices and guide future development to remain consistent with the intended structure of the system \cite{nygard2011documenting}. 

When these decisions are violated intentionally or by oversight, the architecture can slowly drift away from its original design, leading to degradation, increased complexity, and costly maintenance over time \cite{capilla201610}. Detecting such violations early is critical for keeping software architectures aligned with their design intent \cite{karetnikov2024semantic}.
In practice, however, monitoring compliance with ADR is far from straightforward \cite{tofan2013difficulty}. Many decisions are informally documented, scattered across files or repositories, or written in natural language rather than encoded in machine-readable form \cite{buchgeher2023using, alexeeva2016design}. Traditional tools, such as static analyzers or decision-based checkers, often fail to capture this kind of implicit architectural knowledge \cite{bakhtin2022survey, tang2010comparative}. As a result, compliance verification remains a largely manual and error-prone activity, depending heavily on human expertise and contextual understanding \cite{capilla201610}.

Recent advances in Large Language Models (LLMs) have opened new opportunities in this area \cite{dhar2024leveraging, dhar2025draft}. LLMs have demonstrated strong capabilities in understanding both code and natural language, suggesting that they could support higher-level reasoning about software design and architecture \cite{esposito2024leveraging, esposito2024beyond}. Their ability to process heterogeneous information ranging from documentation to source code makes them promising candidates for identifying potential deviations from architectural decision records, even when such decisions are expressed informally \cite{robredo2025EMSE}.

Despite this potential, it is still unclear how reliable these models are when applied to architecture-related tasks \cite{esposito2025large, esposito2025generative}. Most prior work has focused on generating architectural documentation or assisting in decision making \cite{dhar2025draft, arun2025llms, dhar2024leveraging}, while the problem of detecting violations of design decisions has received little empirical attention \cite{moreschini2025ai, karetnikov2024semantic}. Before such models can be integrated into architectural analysis workflows, their accuracy, consistency, and limitations need to be systematically assessed \cite{moreschini2026evolution, esposito2025generative, arun2025llms}.

In this paper, we \textbf{aim} to investigate which LLMs can detect ADR violations in software projects. We analyze 980 ADR extracted from 109 GitHub repositories \cite{buchgeher2023using} and evaluate the performance of multiple models working in a pipeline: one model is responsible for identifying decision violations, while three additional models independently verify its reasoning. 
We measure agreement, accuracy, and precision, and complement the automated results with human expert validation.

Our study shows that LLMs can genuinely help detect when architectural decisions are followed or broken. Across more than a thousand decisions, the models showed substantial agreement, and the best-performing ones achieved accuracy above  90\% in the manually validated sample. They worked well when the decisions were clear and directly visible in the code. However, they struggled when the decisions depended on missing context, infrastructure details, or interactions across different modules. By examining the cases where the models got things wrong, we saw that most errors came from misunderstandings of the decision, missing information, or gaps in technical knowledge. These findings highlight what LLMs can already do and where they still fall short when reasoning about architecture.

This paper makes \textbf{three main contributions}:
\begin{itemize}
    \item A large-scale evaluation of four state-of-the-art LLMs on detecting ADR violations across 109 open-source projects and 980 ADRs.
    \item A multi-model validation pipeline, combining retrieval-augmented detection with independent LLM validators and human expert review.
    \item A taxonomy of failure cases and decision types that reveal where current LLMs struggle, offering actionable insights for architects, researchers, and tool builders.
\end{itemize}

\textbf{Paper structure}:

Section \ref{sec:related} discusses the related works, Section \ref{sec:method} presents the empirical study design, \ReviewerC{Section \ref{sec:result} presents our findings, and Section \ref{sec:discussion} discusses them. Finally, Section \ref{sec:threats} addresses the threats to validity of this work, and Section \ref{sec:conclusion} concludes the paper.}

\section{Related Work}
\label{sec:related}

Research on ADRs has shifted from examining their adoption to understanding their broader impact on software development and architectural practice.

Scheerer et al.~\cite{Scheerer_2017} introduced a method for automatically evaluating complex architectural design decisions. Their approach predicted quality effects, such as performance and reliability, within heterogeneous architectures, showing early promise for supporting architects in industrial design stages.
Therefore, researchers began focusing on preserving architectural knowledge produced during decision-making. 

Shahbazian et al. \cite{Shahbazian_2018} addressed the common issue of poorly documented and easily forgotten decisions. They proposed RecovAr, a technique that automatically recovers architectural design decisions from project artifacts like issue trackers and version histories. Applied to Hadoop and Struts, it reached 75\% recall and 77\% precision, although questions remained about how repository data volume affects accuracy.

Later, Bhat et al. \cite{Bhat2019158} developed Adex, an industrial tool that automatically extracts architectural design decisions from natural-language text to reduce reliance on manual documentation. Adex links decisions to quality attributes and architectural elements, helping prevent architectural erosion. While useful, its deployment in industry requires careful setup and governance.

Buchgeher et al. \cite{buchgeher2023using} mined GitHub projects and found that ADR adoption is still low but steadily increasing. In roughly half the repositories, usage is limited to 1–5 records, while systematic adoption often relies on team-driven processes using templates like Michael Nygard’s. The study also highlights open challenges and opportunities for increasing ADR uptake.

More recently, AI has begun influencing ADR research. Rudra et al. \cite{Rudra_2024} investigated using large language models to generate architectural design decisions. GPT-4 performed reasonably in zero-shot scenarios but was still below human quality; GPT-3.5 and Flan-T5 improved with few-shot learning or fine-tuning. Their findings suggest LLMs can support ADR generation, though more research is needed to reach human-level reliability.
Urdih et al. \cite{Urdih_2025} conducted a literature review on CI/CD pipelines, identifying six architectural design decisions and thirty good practices across 38 grey literature sources. Common techniques include task parallelization, caching, and conditional triggering, with observability enabling further improvements. The study’s main limitation is its reliance on grey literature only.

Existing research shows clear progress but also notable gaps, particularly in detecting violations of architectural design decisions using modern approaches such as large language models.

\section{Empirical study design}
\label{sec:method}
In this Section, we describe the design and execution of our empirical study~\cite{DBLP:books/daglib/0029933}. 
The scripts, the resulting networks, and metrics are available in our Online Appendix and Replication Package~\cite{anonymous_author_2025_17799431}. 

\subsection{Goal and Research Questions}

We formulated the goal of this work as follows:

\textit{Analyze }the use of LLMs \textit{for the purpose of} evaluating their effectiveness in detecting violations of Architectural Decision Records
\textit{with respect to }their agreement, accuracy, and limitations
\textit{from the viewpoint of }software architecture researchers and practitioners
\textit{in the context of }open-source software projects.

To fulfill this goal, we posed the following Research Questions (RQs):

\begin{boxC}{\textbf{RQ$_1$} (Agreement)} 
What is the agreement between the LLMs in detecting Architectural Decision violations?
\end{boxC}

Consistency across models is a critical indicator of reliability. LLMs often differ in their reasoning strategies and sensitivity to prompt phrasing, which can result in different outputs even when given the same input. Measuring the degree of agreement between multiple models sheds light on the stability and reproducibility of their predictions.
High agreement would suggest that LLMs interpret ADRs coherently, whereas low agreement would highlight the need for improved prompt engineering, calibration, or model alignment before practical adoption in architectural analysis.

\begin{boxC}{\textbf{RQ$_2$} (Accuracy)} 
What is the accuracy, precision, and recall of LLMs in detecting Architectural Decision violations?
\end{boxC}

In addition to internal agreement, it is crucial to evaluate the ability of LLMs to accurately and effectively detect genuine decision violations. Performance metrics such as accuracy, precision, and recall provide quantitative evidence of how well these models identify true positives while avoiding false alarms.
This evaluation addresses the effectiveness dimension of the objective, determining whether current LLMs are of sufficient quality to complement or replace traditional static analysis tools. It also allows for comparison between different model families, showcasing their respective strengths and weaknesses in architectural reasoning tasks.

\begin{boxC}{\textbf{RQ$_3$} (Limitations)} 
Which kinds of Architectural Decision and their violations do the LLMs fail to detect reliably?
\end{boxC}

Although LLMs generally perform well, they can struggle with certain categories of ADRs, particularly those involving implicit architectural knowledge, cross-module dependencies, or domain-specific constraints. Recognizing these limitations offers valuable insight into the current capabilities of LLMs and informs future research directions.
Understanding failure patterns also enables the development of hybrid approaches that combine LLM reasoning with static analysis or decision-based techniques, thereby mitigating these weaknesses.

\subsection{Context}
Buchgeher et al.~\cite{buchgeher2023using} provided a dataset of projects that follow and implement ADRs by performing an extensive mining of GitHub projects containing architectural decision records in the form of Markdown plaintext files (Section \ref{sec:related}). At first, the authors identified all GitHub users who have public repositories. Then, they checked if a user has a repository with a Markdown file containing the word ``decision'' and if the full file path also contains ``arch'', ``adr'', ``design'', or ``decision''. Finally, they manually verified the search results and file contents, removing false positives and duplicates to arrive at a set of \textbf{921 repositories} with at least one ADR written to a Markdown file. The contents of the repositories were not otherwise restricted, so the dataset should contain diverse projects written in different programming languages. The authors then analyzed trends among those repositories, such as the amount of committed ADRs, the number of developers working with ADRs, and the timespan during which an ADR is edited. We leverage this dataset as the source of codebases with implemented ADRs.

\subsection{Data Collection}
\label{subsec:datacollection}
In this Section, we describe the data collection process.

\subsubsection{Project Selection}
First, we attempted to clone all the \textbf{921 projects}. Some of the projects were switched to private or deleted after the original dataset creation (January 2022), so we were able to obtain \textbf{882 projects}. After cloning the projects, to prepare them for processing with LLMs, we identified all the files in each project that correspond to source code, i.e., have an extension such as \texttt{*.py} or \texttt{*.java} for the most common programming languages, and aggregated them into a single \texttt{json} file. We checked the MIME type of the files to be \texttt{text/} and converted them to UTF-8 encoding. We took only these source code files and analyzed the distribution and quartiles of TLoC. Based on this analysis, we selected only the projects falling into the 4\textsuperscript{th} quartile, i.e., in our case, having more than 28696 TLoC (Table \ref{tab:quartiles}). Therefore, we reduced our original dataset into \textbf{221 projects}.

We then queried the number of commits of these projects (as of \textbf{2\textsuperscript{nd} of September 2025}). Based on the distribution of these values, we excluded projects in the 1st quartile and thus included only the projects with more than 1184 commits (Table \ref{tab:quartiles}). We were thus left with \textbf{166 projects}.

\subsubsection{ADR Selection}
After selecting the 166 projects, we proceeded to look up the information on the corresponding ADR files from \cite{buchgeher2023using}. The selected projects altogether contain \textbf{1838 ADR files} according to \cite{buchgeher2023using}. 
The authors attempted to determine if the files conform to one of the common Markdown templates for ADR.
Since structured data is easier for the LLMs to process, and we can instruct them how to extract specific information based on the template, we include only ADR for which the template was identified by \cite{buchgeher2023using}. Therefore, we considered \textbf{1235 ADR files} from \textbf{132 projects}.

Moreover, some ADR files might have been moved (renamed) or deleted since the creation of \cite{buchgeher2023using}. We thus checked which of the selected files are still available on the same path as specified in \cite{buchgeher2023using}. Due to this issue, we were left with only \textbf{980 ADR files} that are still available from \textbf{109 projects}.

\begin{boxC}
We obtained a final set of \textbf{980 ADR files} based on availability and parsed templates from \textbf{109 projects}.
\end{boxC}

\begin{table}[]
    \centering
    \caption{Quartiles for project selection: TLoC over all available projects and Commits over 4\textsuperscript{th} quartile of TLoC.}
    \label{tab:quartiles}
    \begin{tabular}{l|rrrrr} \hline 
    \textbf{Metric} & \textbf{Min }& \textbf{Q$_1$} & \textbf{Q$_2$ (Median)} & \textbf{Q$_3$} & \textbf{Max}\\ \hline
    TLoC & 0 & 702.25 & 5218 & 28696.75 &  12,195,341\\ 
    Commits & 1& 1184 & 3694 & 8723& 190,711 \\ \hline 
    \end{tabular}
\end{table}

\subsection{LLM Integration, Design and Validation}
\label{subsec:llmprocess}

This section reports the LLM integration, design, and validation protocol that we adopted from ~\cite{su2025emerging, robredo2025EMSE}.
We assigned a specific role to each model (Table~\ref{tab:model_overview}) based on its responsibility in the data analysis pipeline as follows:

\begin{itemize}
    \item \textbf{Large Reasoning Model (LRM):} Marco-o1 was responsible for extracting the required information and detecting the architectural decision violations. In our study, for each prompt, we received input data (e.g., decisions, code snippets), identified the unique architectural design, and determined whether the source code adhered to this decision, along with a corresponding reason. Its output formed the basis for further verification and was referred to as LRM responses throughout the study.
    \item \textbf{Validation Models (V1/V2/V3):} Mistral-Nemo, Qwen, and Llama were allocated to the validation task. They received the same input as the LRM and were additionally asked to evaluate the LRM's reasoning and output. Their role was to independently assess the correctness of the LRM’s decision (agreement or disagreement).
\end{itemize}

\begin{table*}[hbtp]
  \centering
  \footnotesize
  \caption{Overview of Selected LLMs}
  \begin{tabular}{p{3.5cm} |p{3.5cm} |p{3.5cm} |p{6cm}}
    \hline
    \textbf{Model} & \textbf{Parameters \& Quantization} & \textbf{Details} & \textbf{Highlights} \\
    \hline
    Marco-o1 (LRM) \textsuperscript{a} & 7.6B, not quantized & Inspired by OpenAI’s o-1 & Fine-tuned on CoT datasets, uses MCTS + softmax scoring, excels at math, coding, and logic tasks\\
    
    Mistral-NeMo-Instruct-2407\textsuperscript{b} & 12.2B, not quantized & Fine-tuned version of Mistral-Nemo-Base-2407 & Alignment fine-tuned, supports 128K tokens, outperforms similarly-sized models \\
    
    Qwen3-14B-Base\textsuperscript{c} & 14.8B, not quantized & Decoding Transformer-based & Fine-tuned with enhanced instruction-following, excels in math, programming, and dialogue \\
    
    Llama-3.1-8B-Instruct\textsuperscript{d} & 8B, not quantized & Meta-developed decoding architecture & High performance, strong language understanding and generation, lightweight \\
    \hline

\multicolumn{4}{l}{\textsuperscript{a} \url{https://huggingface.co/AIDC-AI/Marco-o1}, \textsuperscript{b} \url{https://huggingface.co/mistralai/Mistral-Nemo-Instruct-2407}

}\\
   
\multicolumn{4}{l}{\textsuperscript{c} \url{https://huggingface.co/Qwen/Qwen3-14B-Base} ,\textsuperscript{d} \url{https://huggingface.co/meta-llama/Llama-3.1-8B-Instruct}}\\

  \end{tabular}
  \label{tab:model_overview}
\end{table*}

\textbf{Prompting Techniques.} We employed prompt engineering techniques to guide LLMs in the extraction process. According to the state-of-the-art, in-context learning through chat-based prompting provides similar or better results than the more computationally expensive fine-tuning process~\cite{esposito2024beyond}. During the in-context learning phase, each prompt included two components: a \textit{system message}, which established the assistant’s role and specified the expected output format, and a \textit{user message}, which provided the contextual input. The user message contained the current data at hand, e.g., the Decision, Context, and Consequence extracted from the Markdown file. The expected model output consisted of a structured JSON response, and we adopted \textit{Chain of Thought} (CoT) prompting~\cite{wei2022chain}, with few-shot learning.

\textbf{Running LLM.}  To efficiently perform this large-scale analysis, we ran our model using vLLM as a sbatch job on the Mahti supercomputer, hosted by CSC, the Finnish IT Center for Science~\footnote{\url{https://csc.fi}}. Mahti is a high-performance computing system designed for compute- and data-intensive research, featuring over 180,000 CPU cores and a high-speed interconnect network. We utilized up to four NVIDIA A100 GPUs, enabling fast and memory-efficient inference for handling a large volume of data.

\textbf{Human Validation.} Since no previous study reported findings on the accuracy of LLMs for the task at hand, we designed a validation involving three human experts mimicking the LLM validator's roles. The goal is to assess the quality of the model-generated motivations and identify which models consistently produced reasonable outputs.

Our validation followed a three-step protocol. 

\begin{itemize}
    \item One expert reviewed the same input provided to the LRM and the three validation models (V1–V3), and manually evaluated the correctness of each model's motivation. For each case, the expert indicated whether they \texttt{agreed} or \texttt{disagreed} with the LRM’s motivation, noted the majority decision among the validation models, and identified the models they considered correct.
    \item  A second expert repeated the same evaluation independently and documented their level of agreement with the first expert’s judgements. 
    \item  In cases of disagreement between the first two reviewers, a third expert independently assessed the same outputs. Final decisions have been made through majority voting among the three validators.   
\end{itemize}

\subsection{Data Analysis}

In this Section, we report the data analysis process.

\subsubsection{LLM Processing Implementation}

\begin{figure*}[]
    \centering
    \includegraphics[trim={0 0 0 0},clip, width=0.95\linewidth]{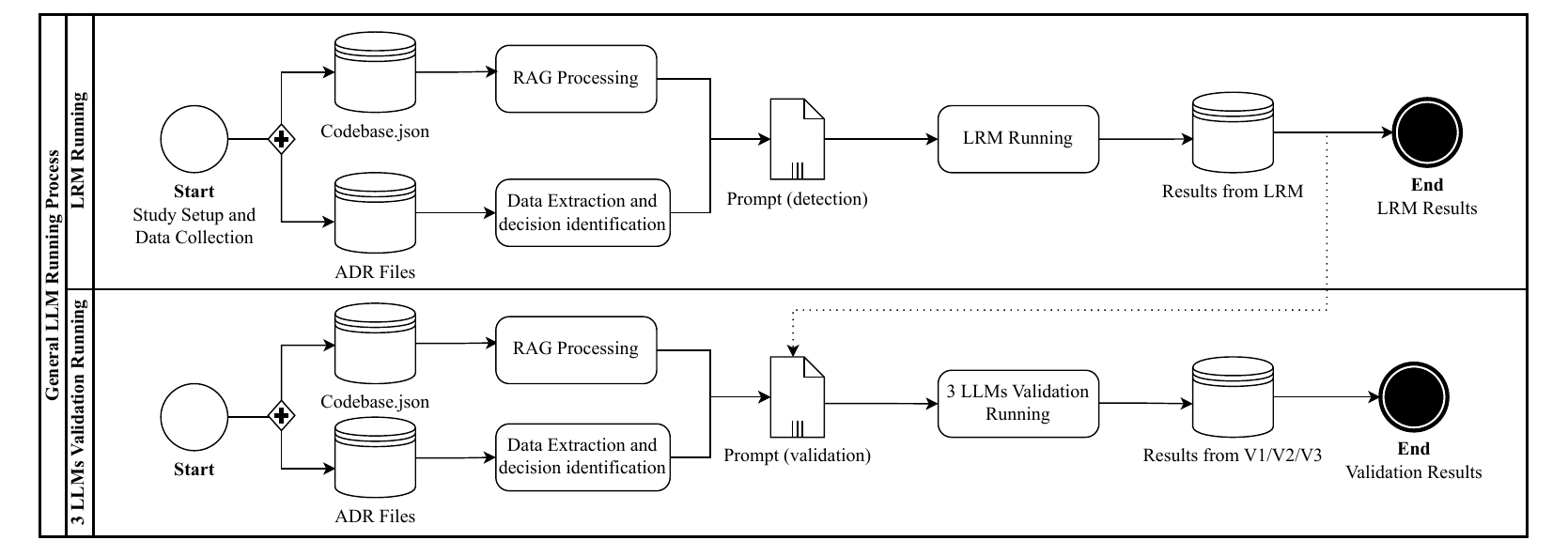}
    \caption{General LLM Processing Implementation}
    \label{fig:llmprocess}
\end{figure*}

We first standardized each repository to match the input format required by the LLM script (Figure~\ref{fig:llmprocess}). Each project was placed in its own folder containing a single JSON file with all source code files and the Markdown files representing its ADRs. We then applied the LLM pipeline described in Section~\ref{subsec:llmprocess} to detect ADR violations. The model received both the ADRs and the source code and was asked to determine whether the code provided sufficient evidence that each decision was implemented or followed.
Because a single Markdown file may contain multiple decisions, we extracted the Decision section and sent it to the model to check whether multiple distinct decisions were present. When this was the case, the script split the file accordingly and evaluated each decision separately. The script also extracted the Status field (e.g., ``Accepted'', ``Proposed'', ``Rejected'') when available, as decision status was treated as a prerequisite in the violation analysis.
Since some JSON files were too large for the LLM’s context window, we adopted a retrieval-augmented generation (RAG) approach. We embedded the entire source code corpus using the ``all-MiniLM-L6-v2'' Sentence Transformer model~\cite{li2020sentence} and indexed the embeddings with a FAISS vector database for efficient nearest-neighbor search~\cite{douze2025faiss}. When performing a semantic query based on the specific ADR, the \textbf{top-k} most similar code snippets (default k=5), based on L2 (Euclidean) distance in embedding space~\cite{johnson2019billion,faiss_metrictypes_wiki}, were retrieved and provided to the model as contextual evidence. RAG was needed to allow the LLM to focus on the most relevant code fragments while avoiding context-length limitations.

Subsequently, we designed the prompt, including: i) Decision, Context, and Consequence parts from the Markdown (ADR) files; ii) the extracted single decision from the Decision part in the ADR; iii) the retrieved code snippets. We instructed LRM to output a JSON object:

\begin{itemize}
    \item \textbf{Decision:} the evaluated single decision from the Decision part in the ADR.
    \item \textbf{Judgement:} ``Compliant'', ``Not Compliant'', or ``Code is Insufficient to Answer''.
    \item \textbf{Reason:} Explain the reason why you have this judgement based on the given contents.
\end{itemize}

In the ``judgment'' filed, we only allowed LRM to return three answers (\textit{hereafter called ``labels''}) as follows:

\begin{itemize}
    \item \textbf{Compliant (C):} The retrieved code snippets provide clear, direct evidence that the decision is implemented or followed as described.
    \item \textbf{Not Compliant (NC):} This applies when: i) The code explicitly contradicts or violates the decision, or ii) There is no relevant evidence that the decision has been implemented (If no code snippet demonstrates any trace of the decision being applied, consider it Not Compliant).
    \item \textbf{Code is Insufficient to Answer (CIA):} This applies when the decision cannot be evaluated based on code, such as organizational or process-related decisions, or cost or other parts that the codebase cannot reflect.
\end{itemize}

We applied the same design principles to both the system message and the user message. The system message defined the LRM’s role as a software architecture expert and outlined the reasoning task, guidelines, clarifications, and output format for determining whether the retrieved code snippets provided sufficient evidence of decision compliance. The user message supplied the task-specific content: the Decision, Context, and Consequence sections of the ADR, the individual decision to be evaluated, and the retrieved code snippets.
To ensure robustness, we integrated several lightweight error-handling mechanisms. These included automatic retries for failed API calls, automatic correction of formatting or incomplete JSON outputs, and continuous logging of processing information. Missing files or sections were recorded explicitly without interrupting execution, allowing processing to continue across projects. Together, these measures improved the reliability and fault tolerance of the evaluation pipeline.

We selected 10 sample projects to test the LRM script, running repeated debugging cycles and manual reviews to refine the prompts and stabilize the pipeline. We executed the finalized script on Mahti, a Finnish supercomputer, to get the complete set of results.
The validation phase followed the methodology in Section~\ref{subsec:llmprocess}. Qwen3, Mistral-Nemo, and Llama3.1 acted as independent validators, functioning as three “experts” assessing the correctness of the LRM’s conclusions. The validation script mirrored the logic of the LRM pipeline, with the models serving exclusively as verifiers. Alongside the original inputs, the script included the LRM’s outputs and used explicit instructions directing each model to judge decision compliance. The system enforced strict formatting and blocked any additional commentary, while the RAG retrieval and error-handling strategies remained aligned with those used in the LRM script. Each validation model was instructed to output a JSON object that determines whether they agree with the results generated by the LRM based on the same input:

\begin{itemize}
    \item \textbf{Judgement:} ``Yes'' or ``No''. 
    \item \textbf{Your answer:} ``C'', ``NC'', or ``CIA''.
    \item \textbf{Exlain:} Reason why you have this judgement and answer based on the given contents.
\end{itemize}

We \textbf{controlled stochastic variability}  by applying consistently across all models the following parameters: temperature=0.1, top-p=0.9, top-k=–1, a repetition penalty of 1.0, and an upper bound of 4096 max new tokens. All models ran through vLLM with an explicitly fixed random seed.

We retrieved the complete output from the LRM and the three validation LLMs. Across the 109 projects and 980 ADR files, the LRM extracted \textbf{1471 individual decisions}. The status fields contained 20 distinct labels, reflecting inconsistent naming conventions across projects. We standardized these labels and mapped them into two categories: ``accepted'' (1) and ``completely rejected'' (0) (Table~\ref{tab:status_mapping}). Applying this mapping allowed us to filter the 1471 decisions and retain only those marked as ``accepted,'' resulting in \textbf{1317 individual decisions}.

\begin{table}[h]
    \centering
    \caption{Mapping of status categories.}
    \label{tab:status_mapping}
    \begin{tabular}{p{7.5cm}|l} \hline
    \textbf{Status} & \textbf{Map} \\ \hline
    Accepted, Proposed, Approved, Implemented, Completed, Decided, WIP, Adopted, Submitted, Proposal & 1 \\ \hline
    Draft, Superseded, Pending, Under review, Rejected, Discussing, Deprecated, Abandoned, DRAFT Not Implemented, not identify any status & 0 \\ \hline
    \end{tabular}
\end{table}

\subsubsection{Agreement in Detecting ADR Violations Among LLMs (RQ$_1$)} We conducted a three-level agreement analysis to evaluate the agreement among LLMs: First, we used \textit{Fleiss' Kappa coefficient} ($\kappa$) \cite{landis1977measurement} to calculate the overall agreement among LLMs. This coefficient measured the degree of agreement among multiple raters while adjusting for random agreement. Second, we calculated agreement by labels, and it reflected the average percentage of original agreement within each label. In contrast to \textit{Fleiss' Kappa}, this metric directly shows the proportion of times the LLMs select the same label without adjusting for chance. Finally, we evaluated the pairwise \textit{Jaccard coefficient} \cite{real1996probabilistic} between any two models to quantify their similarity.

\subsubsection{Accuracy, precision, and recall of LLMs in detecting Architectural Decision violations (RQ$_{2}$)} 

We employed Cocrhan's formula to get a representative sample with a 95\% confidence interval, and a 5\% error margin for human validation \textbf{(sample size = 305)}. Three authors individually manually checked LLM's output for identifying whether: 
\begin{itemize}
    \item the LLMs correctly separated different decisions from the single file; 
    \item the LLMs  correctly interpreted the ADR, and its output was correct based on its understanding.
\end{itemize}
We manually verified all judgments to determine, for each decision, which LLMs produced correct answers. Using our manually verified labels as ground truth, we then quantified the performance of the four LLMs. Across \textbf{305 samples}, we compared the classifications returned by the models (``C/NC/CIA'') against the manual judgments.

We applied standard multi-class classification metrics provided by \textit{scikit-learn} \cite{pedregosa2011scikit, ccarka2022effort}, including overall accuracy, macro-averaged and micro-averaged precision, recall, and F1-score. \textbf{Macro-averaged} precision weighs all classes equally by averaging per-label precision, while \textbf{micro-averaged} precision aggregates predictions across all classes. Therefore, the combined use of macro and micro metrics allows a comprehensive assessment of model performance, capturing both their capability to correctly identify violations across categories and their reliability at the instance level.

We also computed the \textit{Matthews Correlation Coefficient (MCC)} \cite{chicco2020advantages} to evaluate the robustness and balance of predictions across the three classes. MCC directly assesses whether current LLMs demonstrate sufficient accuracy, precision, and recall to detect ADR violations, thereby addressing RQ$_2$.

\subsubsection{Limitations in detecting Architectural Decision and their violations by LLMs (RQ$_{3}$)} We mainly focused on the 305 sample results, where we found that the LLMs provided incorrect answers. 
We compiled all the disagreements (92 among 305) from LLMs. We conducted thematic and axial coding on these 92 cases to identify specific ADR types that LLMs struggle to interpret and the error categories LLMs produce. Two authors independently read these disagreement cases and classified the ADR type and error category. If there is a disagreement, the third author gives the final answer. This approach systematically classifies the reasons that LLMs gave incorrect judgments to analyze the limitations in detecting ADRs and their violations by LLMs.

\section{Results}
\label{sec:result}
In this section, we report the results to answer RQs.

\subsection{LLMs Agreement in Detecting Decisions Violations (RQ$_{1}$)}

\subsubsection{Overall inter-LLMs agreement}

The four LLMs show a high degree of consistency in their ratings that significantly exceeds chance-level agreement (Table~\ref{tab:rq1first}). The Fleiss’ $\kappa$ value of 0.724 shows that their ratings are largely consistent, with only some occasional differences.

\begin{table}[]
\centering
\caption{Overall Inter-LLMs Agreement (Fleiss' Kappa)}
\label{tab:rq1first}
\begin{tabular}{l|rl}
\hline
\textbf{Metric} & \textbf{Value} & \textbf{Interpretation} \\
\hline
Number of items (N) & 1317 & -- \\
Number of raters (n) & 4 & -- \\
Mean observed agreement ($\bar{P}$) & \textbf{0.833} & -- \\
Expected agreement ($P_e$) & \textbf{0.397} & -- \\
\textbf{Fleiss' $\kappa$ (manual)} & \textbf{0.724} & Substantial agreement \\
\textbf{Fleiss' $\kappa$ (statsmodels)} & \textbf{0.724} & Substantial agreement \\
\hline
\end{tabular}
\end{table}

\begin{table}[]
\centering
\caption{Agreement by label with Class Frequency}
\label{tab:rq1second}
\begin{tabular}{l|rr}
\hline
\textbf{Label} & \textbf{Average Agreement} & \textbf{Class Frequency (\%)} \\
\hline
C   & \textbf{0.487} & 58.2 \\
CIA & \textbf{0.157} & 24.7 \\
NC  & \textbf{0.189} & 17.1 \\
\hline
\multicolumn{3}{l}{\textbf{C}: Compliant, \textbf{CIA}: Code is Insufficient to Answer} \\
\multicolumn{3}{l}{\textbf{NC}: Not Compliant} \\
\end{tabular}
\end{table}

\subsubsection{Agreement by labels} 
We computed the average raw proportion of agreement for each label (Compliant ``C'', Not Compliant ``NC'', and Code is Insufficient to Answer ``CIA'') to further investigate how agreement varies for ADR violations across specific labels (Table~\ref{tab:rq1second}). 
The average agreement for label ``C'' is 0.4871, and indicates that the four LLMs reached the same decision in approximately 48.7\% of cases labeled as ``C''. This is the highest agreement among labels, and this class also represents the majority of all ADRs (Class frequency = 58.2\%). This means that when the codebase clearly complies with ADRs, the LLMs tend to demonstrate a stable and convergent interpretation. In contrast, the average agreement for label ``CIA'' is only 0.1574, with a relative frequency of 24.7\%, and the average agreement for label ``NC'' is 0.1893, slightly higher than that of ``CIA'' but still considerably lower than C, and this class represents 17.1\% of the total samples. This limited agreement indicates that identifying clear violations requires less shared context than judging partial inconsistencies, but vague wording or missing justifications can still cause LLMs to diverge.

\begin{figure}[]
    \centering
    \includegraphics[trim={0 0 0 0},clip, width=0.9\linewidth]{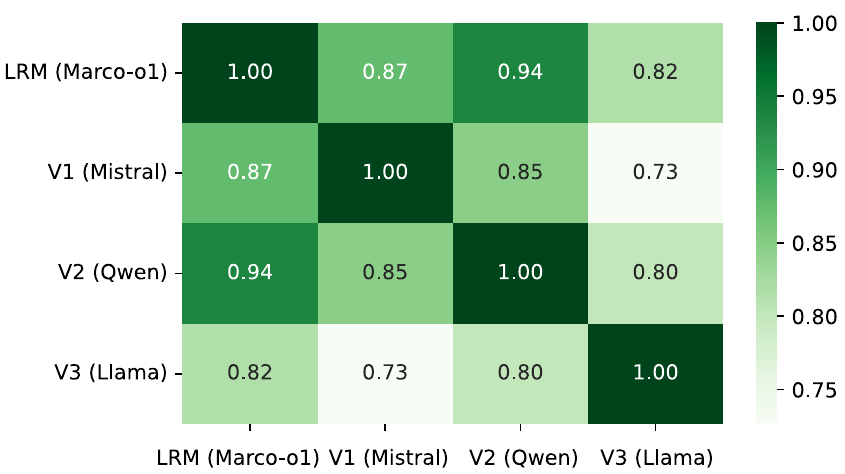}
    \caption{Pairwise Jaccard Coefficients (Overall-C/NC/CIA)}
    \label{fig:jaccardoverall}
\end{figure}

\subsubsection{Pairwise similarity}

The Jaccard coefficients range between 0.73 and 0.94, indicating a generally strong overlap among the four LLMs (Figure~\ref{fig:jaccardoverall}). The highest similarity is between LRM and V2 (Qwen) (0.94), followed by LRM-V1 (0.87) and V1-V2 (0.85). These high values indicate that these three LLMs tend to generate highly consistent predictions and may reflect similar reasoning patterns or training data coverage. In contrast, V3 (Llama) shows relatively lower similarity with the other LLMs ($\approx$ 0.73–0.82), indicating that its architectural or instruction-tuning differences may lead to unique results when detecting ADR violations. This moderate dissimilarity presents the model diversity, despite the overall trends towards consistency.

\begin{figure}[]
    \centering
    \includegraphics[trim={0 0 0 0},clip, width=0.9\linewidth]{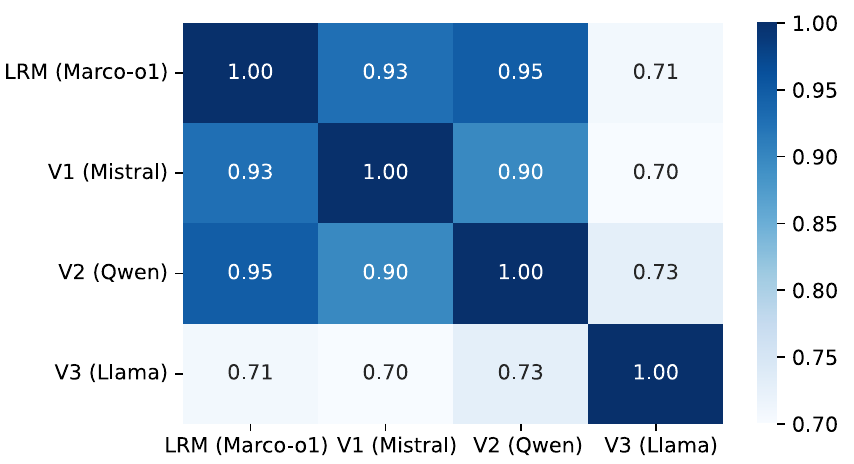}
    \caption{Pairwise Jaccard Coefficients (C)}
    \label{fig:jaccardcompliant}
\end{figure}

We show the Jaccard coefficients restricted to the ``C'' label to further understand the situation of paired consistency in the most common label (Figure~\ref{fig:jaccardcompliant}). We found that the similarity values increase across nearly all model pairs, with most coefficients exceeding 0.90 and reaching up to 0.95 for LRM–V2 (Qwen). This indicates that the LLMs' predictions are highly aligned when LLMs thought the source code followed the ADRs. However, the pairs involving V3 (Llama) remain slightly lower (around 0.70–0.73). In general, LLMs consistently make substantial judgments when detecting ADR violations. However, the slight variation between pairs and the lower similarity involving Llama indicate that different models may differ in their understandings of ADRs.

\begin{boxC}
\textbf{Agreement Analysis (RQ$_{1}$):}
Overall inter-LLMs agreement in detecting ADR violations is \textbf{``Substantial Agreement''}.
LLMs tend to demonstrate a stable and convergent interpretation when the codebase clearly complies with ADRs.
 LLMs generally make strong and consistent judgments when detecting ADR violations. The small differences across model pairs and the lower similarity scores involving Llama suggest that individual models may interpret ADRs differently.
\end{boxC}

\subsection{Performance in Detecting Decisions Violations of LLMs (RQ$_{2}$)}

\subsubsection{Overall performance of four LLMs}

\begin{table}[]
\centering
\scriptsize
\setlength{\tabcolsep}{6pt}
\renewcommand{\arraystretch}{1.2}
\caption{Overall performance comparison of four LLMs}
\label{tab:overall_metrics_transposed}
\begin{tabular}{lcccc}
\hline
\textbf{Metric} & 
\textbf{LRM} &
\textbf{V1 (Mistral Nemo)} &
\textbf{V2 (Qwen3)} &
\textbf{V3 (Llama3.1)} \\
\hline
Accuracy      & 0.911 & 0.852 & 0.904 & 0.829 \\
MCC      & 0.844 & 0.737 & 0.832 & 0.738 \\
P$_{mac}$ & 0.922 & 0.815 & 0.886 & 0.809 \\
R$_{mac}$ & 0.855 & 0.788 & 0.857 & 0.852 \\
F1$_{mac}$ & 0.881 & 0.798 & 0.867 & 0.814 \\
P$_{mic}$ & 0.911 & 0.852 & 0.904 & 0.829 \\
R$_{mic}$ & 0.911 & 0.852 & 0.904 & 0.829 \\
F1$_{mic}$ & 0.911 & 0.852 & 0.904 & 0.829 \\
\hline
\multicolumn{4}{l}{Subscripts: mac = macro average, mic = micro average}
\end{tabular}
\end{table}

We computed the overall performance metrics comparison (Table~\ref{tab:overall_metrics_transposed}), and we found that LRM has the highest accuracy (0.911), followed by V2 (Qwen3) (0.904). While V1 (Mistral Nemo) and V3 (Llama3.1) present relatively lower accuracy (0.852 and 0.829). This indicates that Marco and Qwen can correctly judge more than 90\% of cases, and Mistral and Llama have a higher frequency of misjudgement. Nevertheless, they remain more than 80\% accurate and demonstrate a high level of accuracy. The MCC values also corroborate their accuracy. LRM (0.844) and V2 (Qwen3) (0.832) demonstrate strong predictive ability across classes, while V1 (Mistral Nemo) (0.737) and V3 (Llama3.1) (0.738) demonstrate moderate but less consistent performance. As MCC considers true positives/negatives and false positives/negatives, it indicates that LRM   and V2 (Qwen3) tend to generate more reliable multi-class prediction results.

The macro-averaged precision, recall, and F1-score reflect the unweighted overall performance of three labels (``C'', ``NC'', ``CIA''). We found that LRM   achieves the highest macro-averaged F1 (0.881), and then is V2 (Qwen3) (0.867). This indicates that these two models have a stable performance across all classes. In contrast, V1 (Mistral Nemo) and V3 (Llama3.1) show relatively lower values and indicate relatively unstable performance, especially for minority classes. Interestingly, the micro-averaged scores of all models are almost higher than the macro-averaged ones. This may be because they have weighted each prediction result according to the sample numbers. All four models have the same micro-averaged F1 score (0.911, 0.852, 0.904) as the Accuracy (0.829). This confirms that most predictions come from the dominant class (``C''), and micro averages align closely with overall accuracy. Overall, the results show that LRM   and V2 (Qwen3) outperform V1 (Mistral Nemo) and V3 (Llama3.1) across all metrics, indicating stronger and more reliable judgment behavior when detecting ADR violations.

\subsubsection{Class-wise performance of four LLMs} Table~\ref{tab:classwise_metrics_transposed} reports precision, recall, and F1-scores for each label (C, NC, CIA) of four LLMs.
For the C (Compliant) label cases, all LLMs have very high values, with precision ranging from 0.899 to 0.979 and recall values from 0.7933 to the perfect recall of LRM   (1). LRM, V1 (Mistral Nemo), and V2 (Qwen3) all achieve \text{F1}(C) $>$ 0.9, indicating highly reliable detection of clearly compliant decisions. However, V3 (Llama3.1) exhibits the lowest recall (0.7933), indicating that it misses more true C label cases than the other models.
For the NC (Not Compliant) label, their performance decreases across all models compared to the ``C (Compliant)'' case.  LRM   and V2 (Qwen3) still achieve the strong performace (\text{P}(NC), \text{R}(NC), \text{F1}(NC) > 0.85), While V1 (Mistral Nemo) and V3 (Llama3.1) exhibit weaker results. These indicate that LLMs can be more challenging in correctly identifying ADR violations than in judging compliant ADR, but LRM and Qwen3 still perform well.

The most difficult label is CIA (Code is Insufficient to Answer), and all LLMs have the lowest scores compared to others. LRM   again performs best (\text{F1}(CIA) = 0.792), while Mistral Nemo, Qwen3, and Llama3.1 obtain lower performance (\text{F1}(CIA) = 0.709, 0.759, 0.708). 

All models show low recall for this label, indicating difficulty identifying cases where a decision cannot be evaluated from code alone, such as organizational, process-related, or cost-driven decisions that are not reflected in the codebase.
Across all labels, the models perform best on Compliant (C), moderately on Not Compliant (NC), and weakest on Code Is Insufficient to Answer (CIA). This pattern reflects the increasing complexity of the tasks: detecting violations or missing evidence requires deeper contextual reasoning than confirming compliance. Overall, LRM and Qwen3 continue to exhibit the strongest performance across categories.

\begin{boxC}
\textbf{Accuracy Analysis (RQ$_{2}$):}
Four LLMs indicate a strong and reliable judgment behavior (performance) when detecting ADR violations. However, LRM and Qwen3 are much superior, outperforming Mistral Nemo and Llama3.1 across all metrics.
All LLMs present the best label performance on C, moderate performance on NC, and the poorest performance on CIA.
\end{boxC}

\begin{table}[h]
\centering
\scriptsize
\setlength{\tabcolsep}{6pt}
\renewcommand{\arraystretch}{1.2}

\caption{Class-wise performance comparison of four LLMs}
\label{tab:classwise_metrics_transposed}

\begin{tabular}{lcccc}
\hline
\textbf{Metric} &
\textbf{LRM} &
\textbf{V1 (Mistral Nemo)} &
\textbf{V2 (Qwen3)} &
\textbf{V3 (Llama3.1)} \\
\hline
P(C)      & 0.899 & 0.904 & 0.930 & 0.979 \\
R(C)      & 1 & 0.955 & 0.977 & 0.793 \\
F1(C)     & 0.947 & 0.929 & 0.9537 & 0.876 \\
P(NC)     & 0.934 & 0.746 & 0.8571 & 0.873 \\
R(NC)     & 0.876 & 0.769 & 0.9231 & 0.846 \\
F1(NC)    & 0.904 & 0.757 & 0.8889 & 0.859 \\
P(CIA)    & 0.933 & 0.795 & 0.8723 & 0.577 \\
R(CIA)    & 0.688 & 0.639 & 0.6721 & 0.918 \\
F1(CIA)   & 0.792 & 0.709 & 0.7593 & 0.708 \\
\hline
\end{tabular}
\end{table}

\subsection{Limiations of LLMs in Detecting Decisions Violations (RQ$_{3}$)}
We analyzed 92 cases, out of the 305 manually validated ones (30\%), in which LLMs misjudged ADR compliance. Using thematic and axial coding, we identified the specific types of ADRs that LLMs struggle to interpret and the major failure categories.

\subsubsection{Specific ADR labels that LLMs struggle to interpret}
Table \ref{tab:ADR_categories} summarizes the distribution of ADR labels for which LLMs most frequently produced incorrect judgments.

\begin{table}[h]
\centering
\caption{Specific ADR labels and Distribution}
\label{tab:ADR_categories}
\begin{tabular}{l|rl}
\hline
\textbf{ADR label} & \textbf{\#} & \textbf{Percent} \\
\hline
Infrastructure and Deployment-Specific & 39 & 42.39\% \\
Principle-Driven or Intent-Oriented  & 24 & 26.09\% \\
System and Module Interaction & 16 & 17.4\% \\
Logic or Condition-Intensive & 9 & 9.78\% \\
Context-Dependent  & 4 & 4.35\% \\
\hline
\end{tabular}
\end{table}

\textbf{Infrastructure and Deployment-Specific decisions} (42.39\%) constituted the largest group of errors. These ADRs often involved technologies such as Terraform, Docker, Kubernetes, CI/CD pipelines, or cloud provider-specific configurations, which require specialized knowledge of deployment semantics, environment setup, or infrastructure tools. LLMs misunderstood some implied contents when facing these kinds of decisions, such as the interaction approach of orchestration tools, structural specifications for artifact naming, or deployment directories, leading to wrong or incomplete judgments. LLMs were challenging in reasoning regarding cross-infrastructure technology and environment dependency operations.

\textbf{Principle-Driven or Intent-Oriented decisions} (26.09\%) represented the second largest error source. It reflected architectural guidelines based on design principles or subjective quality objectives (e.g., specification document and test structures). These ADRs were sometimes described broadly and abstractly, leading LLMs to produce ambiguous explanations. This misjudgement often occurs when models attempt to extract specific meanings from decisions that lack clearly defined requirements. This indicated that LLMs struggle to make accurate judgments without explicit operational standards.

\textbf{System and Module Interaction decisions} (17.4\%) reflected the data flow or communication patterns between components, modules, or services. We found that this ADR misjudgement occurred when decisions depended on multi-step workflows, and LLMs had difficulties inferring distributed interactions, system boundaries, or responsibilities between components.

\textbf{Logic or Condition-Intensive decisions} (9.78\%) encoded conditional semantics, branch logic, or version constraints (such as decisions governing release branches or incompatible major versions). LLMs misinterpret these types of decisions by oversimplifying their multi-condition density. This represented the limitations that LLMs process nuanced logical reasoning and interpret complex, multi-part architectural constraints.

\textbf{Context-Dependent decisions} (4.35\%) were the smallest category, yet still important. These decisions relied on contextual conditions or implicit system states, such as resource dependencies or behaviors visible only at runtime. Even when provided with the ADR and the codebase, LLMs struggled because they lack the underlying contextual understanding needed to interpret how such decisions apply in real system scenarios. Although few in number, the recurring difficulties with these decisions underscore the limits of using static, text-based reasoning to evaluate context-sensitive architectural constraints.

\begin{boxC}
\textbf{Limitation Analysis (RQ$_{3}$):}
LLMs struggle the most with infrastructure and deployment-specific ADRs (42.39\% of all errors), indicating that environment-dependent and tool-specific decisions are the hardest for models to interpret.
\end{boxC}

\begin{table}[t]
\centering
\caption{Error Labels and Distribution}
\label{tab:error_categories}
\begin{tabular}{l|rl}
\hline
\textbf{Error label} & \textbf{\#} & \textbf{Percent} \\
\hline
Semantic and Logical Misinterpretation & 41 & 44.57\% \\
Inability to Infer Implicit or Missing Context & 26 & 28.26\% \\
Insufficient Domain and Technical Knowledge & 17 & 18.48\% \\
Overgeneralization and Unsupported Inference & 8 & 8.7\% \\
\hline
\end{tabular}
\end{table}

\subsubsection{Error labels generated by LLMs}
Major failure labels, which indicate the limitations of current LLMs in reasoning tasks related to detecting ADR violations (Table \ref{tab:error_categories}).

\textbf{Semantic and Logical Misinterpretation.} 44.57\% of LLMs' disagreements were due to the misunderstandings of the ADR's logical structure or minor semantic nuances. During the human validation process, we found that sometimes LLMs misjudged partially compliant decisions as fully compliant, and vice versa. This situation indicates the LLMs' limitations, as it is challenging to analyze more detailed logical relationships within ADR texts. When a single ADR contains multiple conditions, optional or implied dependencies, LLMs may simplify or fail to identify all contents, leading to misinterpretations.

\textbf{Inability to Infer Implicit or Missing Context.} There are 28.26\% incorrect predictions caused by the inability to infer implicit or missing context. This means that the key information needed to evaluate ADRs is not fully present in the source code. For instance, some ADRs involve architectural concepts, runtime behaviour, resource consumption, or interactions with external services, as well as code-related aspects. This caused difficulties for LLMs because the lack of contextual information led them to make incorrect judgments.

\textbf{Insufficient Domain and Technical Knowledge.} Although LLMs can understand most concepts and operations in various technologies and domains, their comprehensive knowledge (e.g., system construction, deployment, and infrastructure management frameworks) differs from domain-trained models. For instance, we found that in 18.48\% of the cases, these models made incorrect judgments as they were not able to infer version information or lacked sufficient configuration to evaluate orchestration compliance. Large language models lack a comprehensive understanding of version constraints and configuration scenarios within specific technical domains.

\textbf{Overgeneralization and Unsupported Inference.} The last label is that LLMs may make assumptions and inferences without enough evidence. In our samples, we found a few cases (8.7\%) where LLMs made assumptions or inferences not explicitly present in the source code. These incorrect results reflect a typical behavior of LLMs that they tend to overgeneralize and even express their own opinions based on assumptions rather than express uncertainty. This kind of answer deviation may potentially occur, although we have repeatedly emphasized in the prompt to prevent this situation.

\begin{boxC}
\textbf{Error labels generated by LLMs (RQ$_{3}$):}
LLMs fail primarily due to semantic and logical misinterpretation (44.57\% of errors), followed by missing-context inference issues (28.26\%), domain-knowledge gaps (18.48\%), and occasional overgeneralization (8.7\%).
\end{boxC}

\section{Discussion}
\label{sec:discussion}
This section further discusses the main results achieved in our study and reports some lessons learned and their implications for researchers and practitioners. 

\textbf{Reliability of LLM Judgements Through Inter-Model Agreement.} The agreement analysis shows that LLMs achieve substantial alignment when judging ADR compliance or violations. This is notable given the ambiguity of natural-language ADRs and the diversity of architectural decisions across projects. The high overall agreement (83.38\%) suggests that the models often converge on similar interpretations of decisions and code evidence, even without formal specifications.
However, agreement varies across categories. The strongest consistency appears in Compliant (C) cases, where clear code implementations allow all models to reach stable conclusions. In contrast, agreement drops for Not Compliant (NC) and Code Is Insufficient to Answer (CIA), where detecting missing or indirect evidence requires deeper reasoning, an area where current LLMs remain less reliable.
Pairwise similarity results reinforce this trend. Marco-o1 and Qwen3 show high overlap, indicating similar reasoning patterns or stronger sensitivity to ADR structure. Llama diverges more frequently, suggesting reduced stability when architectural reasoning becomes less explicit. While LLMs can function as useful ``second readers'' for ADRs, model choice matters, and some models align more naturally with architectural reasoning tasks than others.

\textbf{Strengths and Weaknesses in Architectural Violation Detection.} The performance results confirm that LLMs, especially Marco-o1 and Qwen3, achieve strong accuracy, with both models exceeding 90\% correctness in the manually validated subset. This shows that LLMs can meaningfully support the detection of architectural decision violations in real-world repositories when combined with well-designed prompts and retrieval-augmented context.
However, the class-wise analysis reveals clear limits. All models perform best on Compliant cases, indicating that confirming alignment with a decision is far easier than detecting violations. Accuracy drops for Not Compliant cases, reflecting the greater difficulty of recognizing missing or contradictory behavior. The weakest results appear in the CIA category, where models must recognize that the decision cannot be evaluated from the available code. This requires meta-reasoning about evidence completeness, an ability LLMs still struggle with.
Precision and recall patterns reinforce these findings. Marco-o1, for example, shows perfect recall for Compliant decisions, demonstrating its strength in identifying positive evidence. Yet recall decreases for CIA cases, indicating difficulty distinguishing between ``no violation'' and ``insufficient information''. This underscores a broader challenge: separating actual violations from missing evidence is subtle and often requires architectural judgment that goes beyond simple pattern matching.

\textbf{Implications for Architectural Practice.} Overall, the results indicate that LLMs can already offer meaningful support for architectural compliance checking, especially for ADRs that are clearly expressed and have direct counterparts in the code. In such cases, LLMs can help automate routine reviews, highlight potential violations, and assist architects in navigating large and complex repositories.
However, they are not yet reliable enough to replace human oversight. Architectural reasoning often depends on implicit constraints, historical context, and system-level behavior that cannot be inferred from code snippets alone. The uneven performance across categories and models shows that LLM-based compliance checking is best used in hybrid workflows, where model outputs serve as guidance or early warnings rather than final decisions.
The results also highlight the importance of retrieval quality. Since the pipeline depends heavily on the RAG step, any limitations in retrieved evidence directly affect the model’s reasoning. Improving retrieval precision and context reconstruction may offer greater benefits than further prompt refinement.

\textbf{Toward More Mature LLM-Driven Architectural Analysis.} The results point to several promising directions for future work. Standardizing ADR formats could improve consistency in how LLMs interpret decisions and reduce ambiguity. Combining static analysis with LLM-based reasoning may help detect more subtle violations that depend on detailed dependency or execution-flow understanding. In addition, model calibration and cross-model ensembles could enhance reliability by leveraging complementary strengths across different LLMs.
Overall, while current models already provide useful insights, achieving dependable large-scale ADR violation detection will require further improvements in retrieval quality, prompt design, and hybrid human-in-the-loop workflows.

\section{Threats to Validity}
\label{sec:threats}

\textbf{Construct Validity.} We used the dataset from Buchegeher et al.~\cite{buchgeher2023using}, which means our selection process may be biased by their data-collection process. Their approach began by identifying all GitHub users with Markdown files and then filtering to repositories containing ADRs. Despite this, our analysis of TLoC and commits showed a diverse mix of large and small projects, many actively maintained by organizational accounts.
We further limited our study to ADRs that could be parsed with a known template, which excludes potentially valuable records that did not match the template. Some repositories, or specific files within them, were no longer available under the names or paths reported in \cite{buchgeher2023using}. As a result, we may have missed ADRs that were simply renamed.
A more robust approach would involve re-mining all Markdown files from the selected repositories or tracking ADR file changes directly through version control since the original snapshot~\cite{buchgeher2023using}.

\textbf{Internal Validity.} The study can be influenced by some factors. The labeling of violations depends on the human interpretation of ADR and the corresponding source code, which introduces subjectivity and ambiguity. We selected the LLM approach to identify architectural design violations from the dataset of Buchegeher et al.~\cite{buchgeher2023using}. The LLM approach can have problems during generation, such as hallucinations. To properly assess this method, we performed validation with three different LLM models and human validation on the generated results to ensure correctness. 

\textbf{External Validity.} The dataset and context of the analyzed projects constrain the generalizability of the findings. The study is performed on open-source GitHub repositories that include ADR. The conclusions may not be directly generalized to industrial or other fields. The dataset includes popular programming languages and actively maintained projects, which may not reflect the diversity of software ecosystems.

\textbf{Conclusion Validity.} The main conclusion threat relates to the manual validation of the 305 sampled cases, which serves as the ground truth for accuracy, precision, recall, and other metrics. Even though multiple reviewers participated and disagreements were resolved through majority voting, human interpretation of ADRs and code snippets may still introduce inconsistencies. A second threat is the category imbalance, which may inflate overall accuracy and micro-averaged results. To reduce this risk, we also reported macro-averaged metrics and MCC, though some sensitivity to class distribution remains. Finally, the conclusions assume that the sampled cases are representative.

\section{Conclusion}
\label{sec:conclusion}
We investigated how LLMs detect violations of ADR across 109 open-source projects. Our multi-model evaluation shows that LLMs can reliably identify many decision violations when ADRs are explicit and directly reflected in code. Agreement across models is substantial, and overall accuracy exceeds 90\% for the strongest performers.
However, the models struggle with decions tied to infrastructure, implicit intentions, multi-module interactions, or context that is not visible in code. The CIA label remains the hardest to judge, revealing limits in distinguishing ``no evidence'' from ``non-compliance'.' Errors typically stem from semantic misinterpretation, missing context, and incomplete technical knowledge.
LLMs can meaningfully support architectural compliance checks, but cannot replace human reasoning or complementary analysis tools.

\section*{Declaration on the use of generative AI}
The authors used ChatGPT for suggestions on improving textual clarity. All research design, data analysis, interpretations, and manuscripts were created by the authors themselves.

\balance
\bibliographystyle{IEEEtran}
\bibliography{main}

\end{document}